\documentclass[aps,prl,twocolumn,superscriptaddress]{revtex4}

\usepackage{graphicx} 
\usepackage{bm} 
\usepackage{multirow} 
\usepackage{subfigure}
\usepackage{amsfonts}
\usepackage{epstopdf}
\usepackage{color}

\begin{document}

\title{Defect formation dynamics during CdTe overlayer growth}

\author{J. J. Chavez}
\affiliation{University of Texas at El Paso, El Paso, TX. 79968, USA}

\author{D. K. Ward}
\affiliation{Sandia National Laboratories, Livermore, California 94550, USA}

\author{B. M. Wong}
\affiliation{Sandia National Laboratories, Livermore, California 94550, USA}

\author{F. P. Doty}
\affiliation{Sandia National Laboratories, Livermore, California 94550, USA}

\author{J. L. Cruz-Campa}
\affiliation{Sandia National Laboratories, Albuquerque, NM 87185, USA}

\author{G. N. Nielson}
\affiliation{Sandia National Laboratories, Albuquerque, NM 87185, USA}

\author{V. P. Gupta}
\affiliation{Sandia National Laboratories, Albuquerque, NM 87185, USA}

\author{D. Zubia}
\affiliation{University of Texas at El Paso, El Paso, TX. 79968, USA}

\author{J. McClure}
\affiliation{University of Texas at El Paso, El Paso, TX. 79968, USA}

\author{X. W. Zhou}
\email[]{X. W. Zhou: xzhou@sandia.gov}
\affiliation{Sandia National Laboratories, Livermore, California 94550, USA}

\date{\today}

\begin{abstract}

The presence of atomic-scale defects at multilayer interfaces significantly degrades performance in CdTe-based photovoltaic technologies. The ability to accurately predict and understand defect formation mechanisms during overlayer growth is, therefore, a rational approach for improving the efficiencies of CdTe materials. In this work, we utilize a recently developed CdTe bond-order potential (BOP) to enable accurate molecular dynamics (MD) simulations for predicting defect formation during multilayer growth. A detailed comparison of our MD simulations to high-resolution transmission electron microscopy experiments verifies the accuracy and predictive power of our approach. Our simulations further indicate that island growth can reduce the lattice mismatch induced defects. These results highlight the use of predictive MD simulations to gain new insight on defect reduction in CdTe overlayers, which directly addresses efforts to improve these materials.

\end{abstract}



\maketitle


The cost of electrical energy generated using CdTe-based multilayer solar cells has reached \$0.15/kWh, lower than any other photovoltaic technology \cite{Z2010}. These CdTe-based solar cells can profoundly change energy supplies if the 17.3\% energy efficiency achieved today is significantly improved towards the 29\% theoretical value \cite{STTWK1999,DVHC2000}. The current inefficiency of the CdTe solar cells is attributed to charge-trapping defects at the multilayer interfaces \cite{BCCCKLSJ2007,ZJWZZ2008,STYLBFJ2001}. A recently developed CdTe bond-order potential (BOP) \cite{WZWDZP2012,ZWWD2012} has enabled molecular dynamics (MD) simulations of defect formation to approach a quantum-mechanical accuracy level. The objective here is to perform such MD simulations to explore defect formation during vapor deposition of CdTe overlayers. These simulations provide critical insight required to improve the energy efficiency of CdTe modules.

MD simulations of semiconductor vapor deposition are extremely challenging because they sample a large number of metastable configurations not known a priori. If the interatomic potential used in a simulation over-predicts the cohesive energy magnitude of any of these configurations, that configuration will likely persist, resulting in an unrealistic amorphous film that offers no useful information. A vast majority of previous MD simulations of semiconductor vapor deposition \cite{GFBTBR1997,XGRSBCFKSMC1994,GHR1998,SR1995,LF2000} were achieved using Stillinger-Weber (SW) \cite{SW1985} potentials. It has been established \cite{WZWDZ2011} that while SW potentials can easily ensure crystalline growth, they cannot satisfactorily capture the property trends of other configurations and, hence, they cannot accurately reveal defect formation. Tersoff potentials \cite{T1989}, on the other hand, can capture property trends more accurately. However, this also makes Tersoff potential difficult to parameterize to ensure the lowest energy for the equilibrium phase \cite{WZWDZ2011}. As a result of not capturing the lowest energy phase due to poor parameterization, many literature Tersoff potentials \cite{OG1998,NFOTMO2000,AJCHW1995} incorrectly predict amorphous growth. Not surprisingly, we found \cite{WZWDZ2011} that the existing CdTe Stillinger-Weber \cite{WSM1989} and Tersoff \cite{OG1998} types of potentials have not sufficiently addressed issues involving defects. The CdTe BOP \cite{WZWDZP2012,ZWWD2012} makes a significant stride towards improving semiconductor simulations because (a) it is analytically derived from quantum mechanical theories and its quantum accuracy has been widely documented \cite{PFNMZW2004,PO1999,PO2002,DMNZWP2005}; (b) it goes beyond Tersoff potentials on transferability and is well parameterized to capture properties of a large number of elemental and compound configurations spanning coordination of 1 to 12 including small clusters, bulk lattices, and defects; and most importantly, (c) it predicts crystalline growth \cite{ZWWD2012}.

\begin{figure}
\includegraphics[width=3.3in]{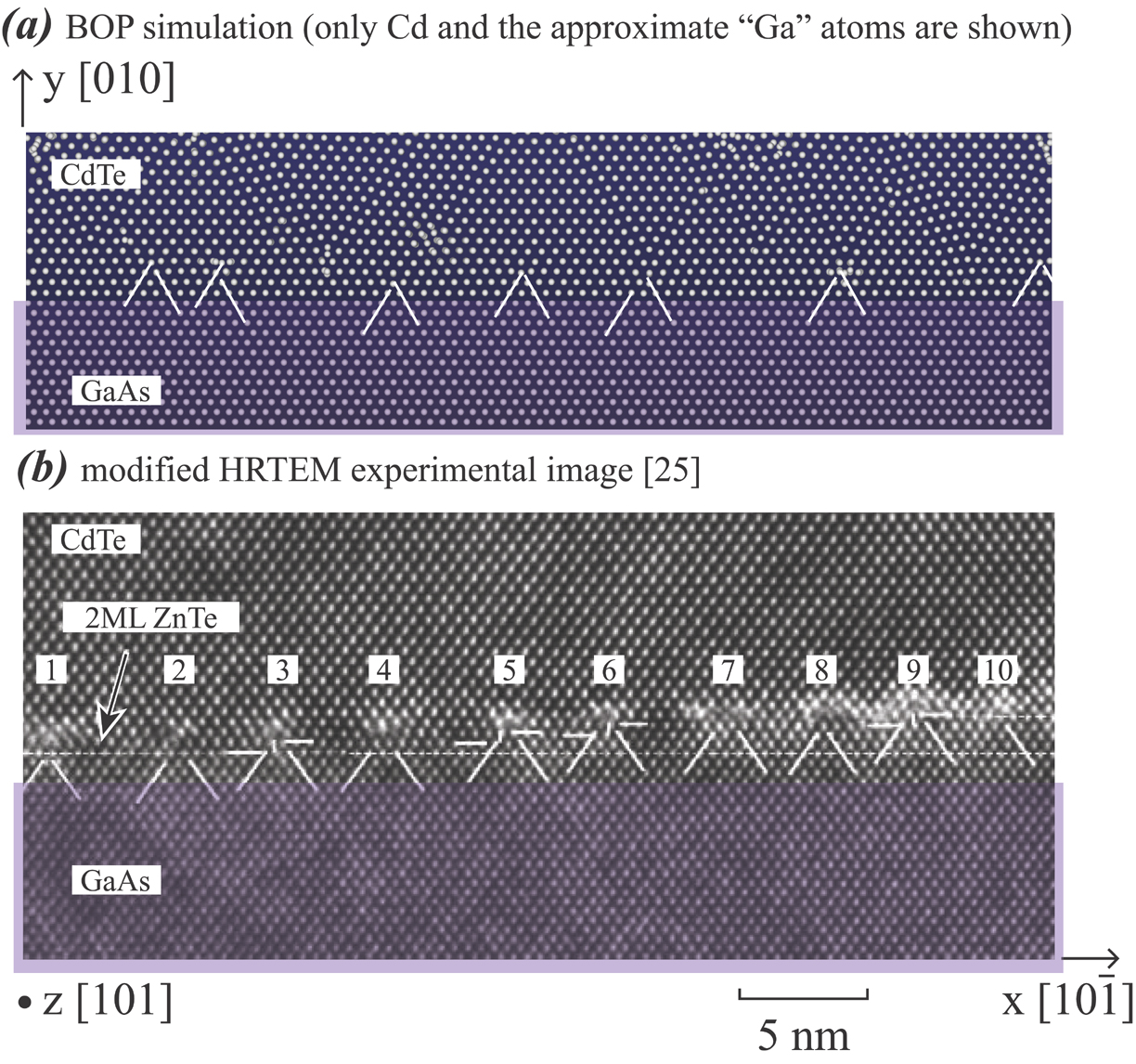}
\caption{BOP simulation and HRTEM image \cite{KDDL2003} of atomic structure of CdTe-on-GaAs multilayers.
\label{misfit}}
\end{figure}

High-resolution transmission electron microscopy (HRTEM) experiments have been performed to examine defects in CdTe/GaAs multilayers with a lattice mismatch of $\epsilon_0 \approx$ 12.78\% \cite{KDDL2003}. To directly compare with the experiments, we performed an MD simulation of CdTe overlayer growth using the same lattice mismatch. The computational system, shown in Fig. \ref{misfit}(a), is periodic in the x- and z- directions containing 100 ($10\bar{1}$) and 8 ($101$) planes, respectively. To incorporate the lattice mismatch with only the CdTe BOP, a substrate containing 35 ($040$) planes in the y- (thickness) direction was compressed by 12.78\% in the x- dimension to match the size of GaAs. To prevent the dimension from relaxing back to that of CdTe, the atomic positions of the bottom 25 ($040$) planes were fixed during a constant volume MD vapor deposition simulation. An adatom incident kinetic energy of 0.1 eV, an incident direction normal to the surface, a substrate temperature of 1000 K, a stoichiometric vapor ratio of Cd/Te = 1, and a deposition rate of around 0.96 nm/ns were used. MD simulations of vapor deposition must be performed at accelerated deposition rates due to the computational cost. While this may lead to overestimates of kinetically-trapped defects such as vacancies, it conservatively (and hence correctly) reveals the formation of non-kinetically-trapped defects such as misfit dislocations. The configuration obtained after about 4 ns of deposition is shown in Fig. \ref{misfit}(a) with comparison to a modified HRTEM image \cite{KDDL2003} shown in Fig. \ref{misfit}(b).

Fig. \ref{misfit}(a) indicates 7 misfit dislocations near the interface. These dislocations are clearly the edge type of Lomer dislocations with two extra planes about 144.7$^o$ from the y- axis. Both the dislocation configurations and average dislocation spacing are in remarkably good agreement with the experimental results shown in Fig. \ref{misfit}(b).

\begin{figure}
\includegraphics[width=3.3in]{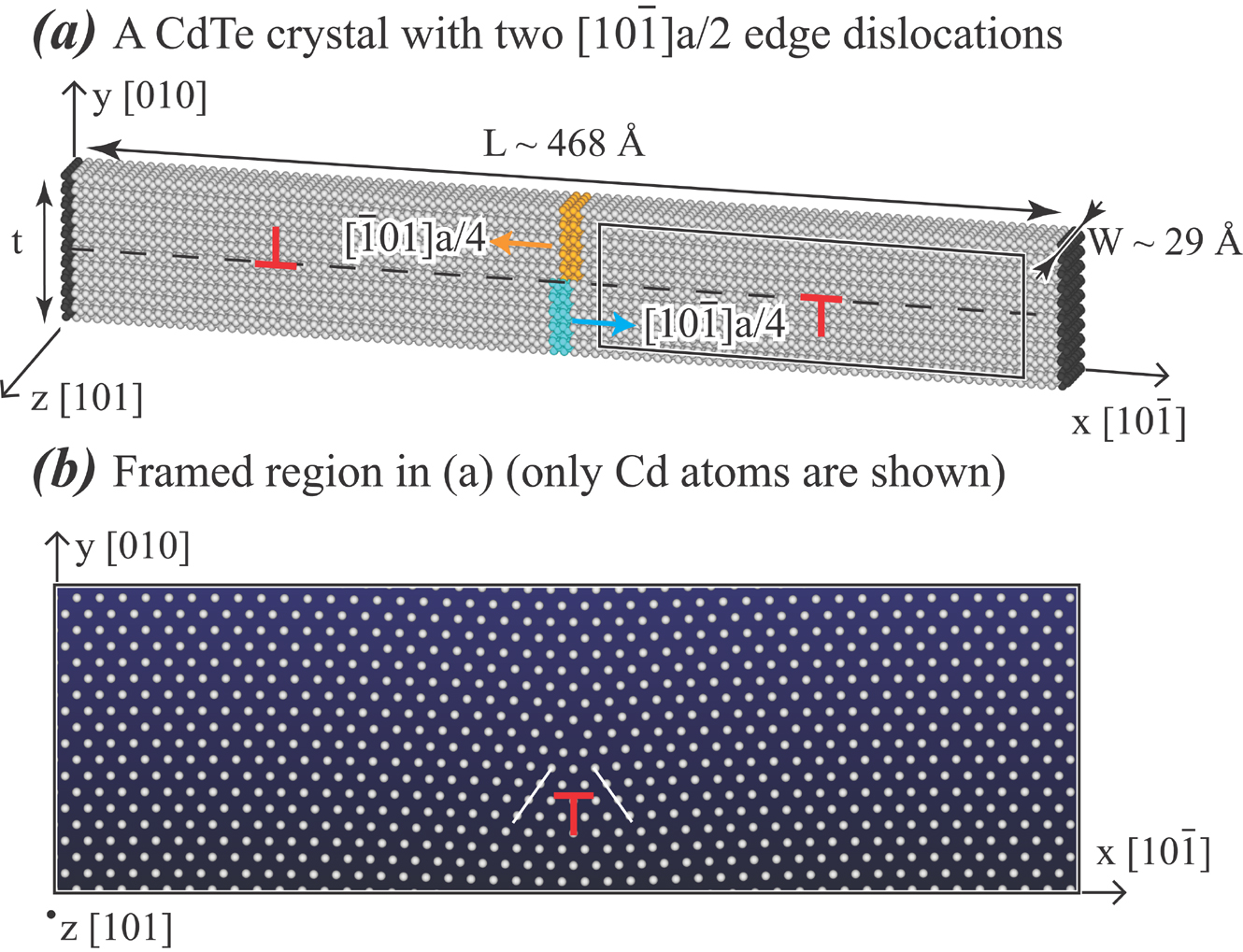}
\caption{Relaxed [$10\bar{1}$]a/2 misfit dislocation.
\label{model}}
\end{figure}

Analysis of simulated results indicates that the misfit dislocations have a [$101$] line direction and a [$10\bar{1}$]a/2 Burgers vector. To verify this, a dislocation model was developed in Fig. \ref{model}(a). With an original offset of a Burgers vector, the middle orange region in the upper half and the middle blue region in the lower half of a CdTe crystal are pushed to the left and the right respectively by a half of the Burgers vector until they are aligned (the displacements of other atoms are ramped under the condition that the black regions at the ends of the system remain unchanged). Relaxed dislocation configurations were determined using an energy minimization simulation with the orange and blue regions held as a rigid body (the black regions are also treated as rigid bodies). This process results in two relaxed, symmetric dislocations of opposite sign. To verify the dislocation model, one of the dislocation configurations created in the framed region shown in Fig. \ref{model}(a) is examined in Fig. \ref{model}(b). It can be seen that dislocations created in Fig. \ref{model}(a) match precisely with the configurations seen from Figs. \ref{misfit}(a) and \ref{misfit}(b). Because the edge dislocation model shown in Fig. \ref{model}(a) does not insert or remove atoms, it can be referenced to a dislocation-free system with the same number of atoms to calculate dislocation energy \cite{ZWWDZ2012}, as will be discussed below. 

Using the model shown in Fig. \ref{model} to create an initial dislocation configuration containing 153 atoms, both density function theory (DFT) and BOP methods were used to calculate the relaxed dislocation core structures. We found that the core structures obtained from the two methods are similar \cite{supple1}.
 
\begin{figure}
\includegraphics[width=3.3in]{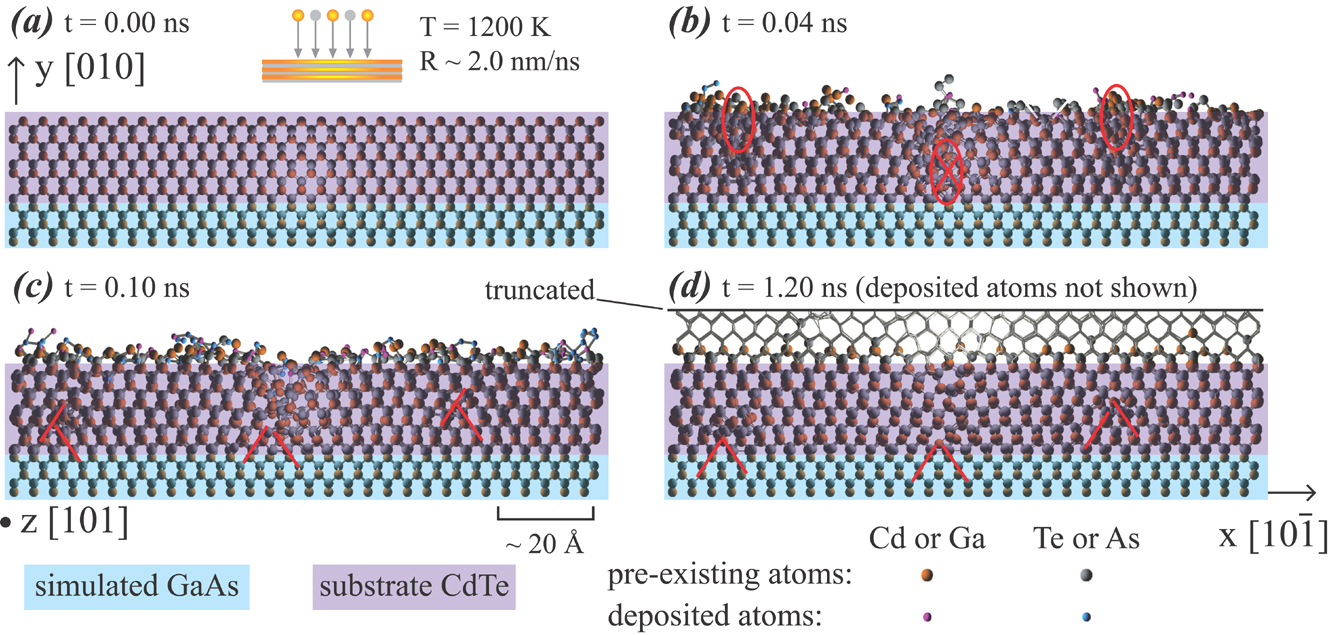}
\caption{Time resolved film configurations. To clearly reveal the diffusion of pre-existing atoms, the deposited atoms are not directly shown in (d) but are displayed at the intersections of the bars.
\label{mechanism}}
\end{figure}

To examine misfit dislocation formation mechanisms, time resolved configurations obtained from a smaller scale, similar vapor deposition simulation are examined in Figs. \ref{mechanism}(a)-(d), where the blue and purple regions are respectively the GaAs underlayer and the pre-existing CdTe substrate prior to the deposition. No dislocations exist in the CdTe film at the start of simulation shown in Fig. \ref{mechanism}(a). Three distorted surface regions (marked by the red ellipses) are seen in Fig. \ref{mechanism}(b) at time 0.04 ns. These distorted regions correspond to the [$10\bar{1}$]a/2 misfit dislocations as exemplified by the middle ellipse where two extra planes indicated by the red lines have emerged. The extra planes become clear after 0.10 ns as shown in Figs. \ref{mechanism}(c) and \ref{mechanism}(d). A comparison between Figs. \ref{mechanism}(c) and \ref{mechanism}(d) clearly indicates that upon nucleation at the surface, the misfit dislocation cores (locations where the extra planes terminate) continuously climb towards the interface in an approximate vertical direction. The climbing mechanism is further verified in Fig. \ref{mechanism}(d) as the pre-existing atoms are seen to extensively diffuse out of the upper boundary of the purple region.

Unlike kinetically-trapped defects, the predicted dislocations are realistic at the accelerated deposition rate (i.e., more dislocations are expected should the deposition be reduced). The equilibrium dislocation densities can be calculated from the misfit strain energy density and dislocation line energy \cite{PNLC1993,N1989}. Here the volume density of misfit strain energy $e_s$ is expressed as 
\begin{equation}
e_s = C \cdot \epsilon^2 \cdot \left(\alpha+t\right) / t 
\label{strain energy}
\end{equation}
where $\epsilon$ is the strain, t is film thickness, and $C$ and $\alpha$ are constants. The parameter $\alpha$ provides a small adjustment to the thickness to account for the surface effect. By creating strained and unstrained CdTe films at different thicknesses, the strain energy densities were calculated using the BOP-based energy minimization simulations. Fitting the results to Eq. (\ref{strain energy}) yields $\alpha$ = -3.723 $\AA$, and $C$ = 0.307 $eV/\AA^3$.

\begin{figure}
\includegraphics[width=3.3in]{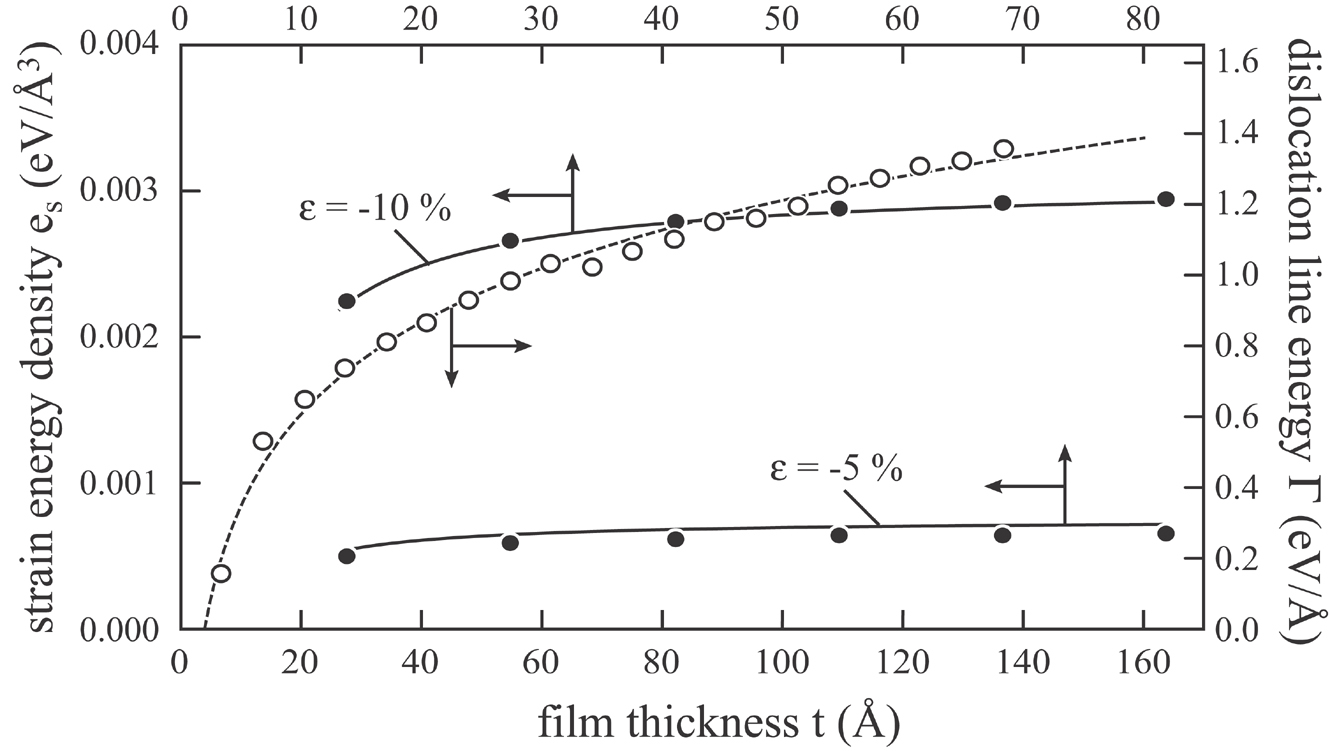}
\caption{Strain and dislocation line energies.
\label{energies}}
\end{figure}

The dislocation line energy $\Gamma$ can be written as \cite{PNLC1993}
\begin{equation}
\Gamma = A + B \cdot ln\left(t\right)
\label{dislocation line energy}
\end{equation}
where A and B are constants. By creating dislocated [using the model shown in Fig. \ref{model}(a)] and perfect CdTe films at different thicknesses, the dislocation energies were calculated using energy minimization simulations. Fitting the results to Eq. (\ref{dislocation line energy}) yields A = -0.520 $eV/\AA$, and B = 0.376 $eV/\AA$. The fitted and simulated strain energy densities and dislocation line energies are all shown in Fig. \ref{energies} as a function of film thickness. It can be seen that the fitted relations well represent the simulations.

Applying Eqs. (\ref{strain energy}) and (\ref{dislocation line energy}) in the classic misfit dislocation theory \cite{PNLC1993,N1989}, we found a critical film thickness for dislocation formation, $t_c$ = 3.7 $\AA$, and an equilibrium dislocation spacing in a thick film, $d \approx$ 38 $\AA$ (corresponding to 10 - 11 dislocations in Fig. \ref{misfit}). Both a small critical thickness and the calculated dislocation spacing match well Figs. \ref{misfit} and \ref{mechanism}. These BOP calculations significantly improve over the continuum theories because they accurately capture dislocation core energies. In addition to analyzing static properties of dislocations, the BOP method is also effective in quantifying dislocation mobility \cite{ZWWDZ2012}.

\begin{figure}
\includegraphics[width=3.3in]{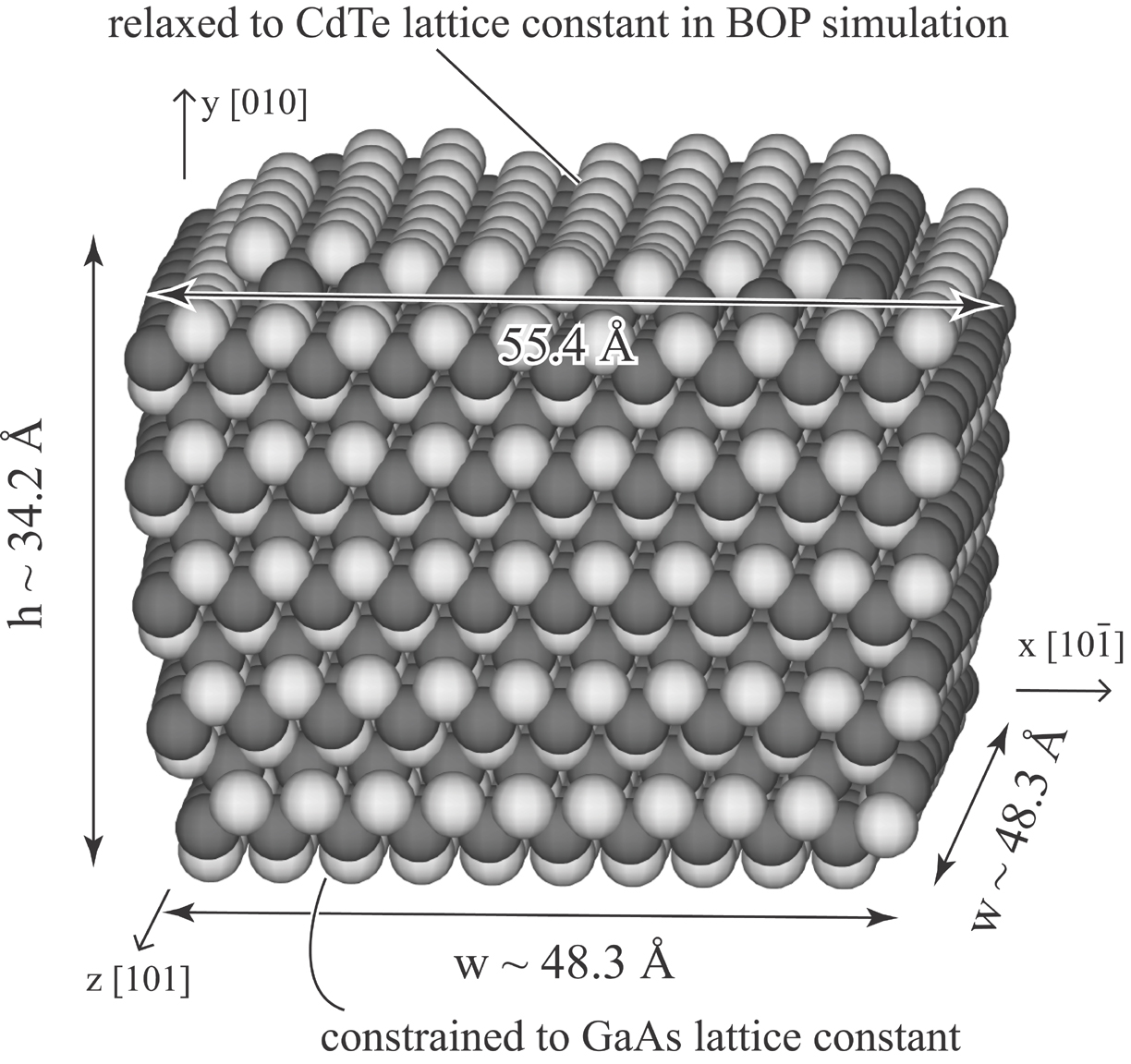}
\caption{Three dimensional relaxation of a CdTe island.
\label{island}}
\end{figure}

The discoveries of an extremely high misfit dislocation density even using accelerated deposition, and the direct surface nucleation of misfit dislocation at a very small critical film thickness, indicate that for continuous multilayered films, these defects cannot be reduced kinetically, or by improving the quality of the substrate. To explore novel methods to reduce defects, an isolated CdTe island containing 10 ($10\bar{1}$), 5 ($010$), and 10 ($101$) planes in the x-, y-, and z- directions respectively is studied. A BOP-based energy minimization was used to relax the island with the bottom two ($040$) planes constrained to the GaAs lattice dimension and the remaining part free to move. The final configuration obtained from the simulation is shown in Fig. \ref{island}. Fig. \ref{island} indicates that for a small island (width around 48.3 \AA), the large lattice mismatch between CdTe and GaAs (as simulated by the bottom part of the island) is completely relaxed over a height distance of about 34.2 \AA~due to the three dimensional relaxation of the island. This means that island growth can reduce the lattice mismatch induced defects. We are currently combining BOP simulations and experiments to further explore this nanoscale phenomenon.

In conclusion, a BOP-based method is shown to approach a quantum-mechanical fidelity capable of predicting crystalline growth and misfit dislocation formation during extremely challenging MD vapor deposition simulations of semiconductor multilayers. The predicted misfit dislocation configuration and density in the CdTe/GaAs multilayers are seen to accurately match the HRTEM experiments. This is a significant improvement over the previous models, and provides a powerful theoretical tool to study defect formation in important materials systems. The BOP simulations also reveal a very small critical film thickness of 3.7 \AA, and surface nucleation and climb mechanisms of misfit dislocation formation. These results strongly indicate that continuous CdTe/GaAs multilayers always contain a very high misfit dislocation density regardless growth conditions. The discovery that the CdTe/GaAs misfit strain can be completely relaxed in a CdTe island over a short distance indicates that it is still possible to create dislocation-free CdTe overlayers using nano-patterned island growth. To guide specific experiments to explore such a possibility, we are currently using the BOP simulations to develop an analytical relation between dislocation density and island size. 

\begin{acknowledgments}

This work is supported by the DOE/NNSA Office of Nonproliferation Research and Development, Proliferation Detection Program, Advanced Materials Portfolio, and The National Institute for Nano-Engineering (NINE). Sandia National Laboratories is a multi-program laboratory managed and operated by Sandia Corporation, a wholly owned subsidiary of Lockheed Martin Corporation, for the U.S. Department of Energy's National Nuclear Security Administration under contract DE-AC04-94AL85000.

\end{acknowledgments}

\appendix

\end{document}